\documentclass[twocolumn]{aastex63}

\usepackage{booktabs}
\usepackage{array}
\usepackage{multirow}
\usepackage{nameref}
\usepackage[T1]{fontenc}
\received{November 30, 2020}
\revised{February 16, 2021}
\accepted{February 19, 2021}
\submitjournal{ApJ}

\newcommand{\p}[1]{{\color{magenta}{#1}}}

\newcommand{\eg}{{e.g.},}

\newcommand{\cf}{{cf.}}

\shortauthors{To et al.}
\graphicspath{{./}{figures/}}
\begin{document}

\title{The Evolution of Plasma Composition During a Solar Flare}

\author[0000-0003-0774-9084]{Andy S.H. To}
\affiliation{University College London, Mullard Space Science Laboratory, Holmbury St. Mary, Dorking, Surrey, RH5 6NT, UK}

\author[0000-0003-3137-0277]{David M. Long}
\affiliation{University College London, Mullard Space Science Laboratory, Holmbury St. Mary, Dorking, Surrey, RH5 6NT, UK}

\author[0000-0002-0665-2355]{Deborah Baker}
\affiliation{University College London, Mullard Space Science Laboratory, Holmbury St. Mary, Dorking, Surrey, RH5 6NT, UK}

\author[0000-0002-2189-9313]{David H. Brooks}
\affiliation{College of Science, George Mason University, 4400 University Drive, Fairfax, VA 22030, USA}

\author[0000-0002-2943-5978]{Lidia van Driel-Gesztelyi}
\affiliation{University College London, Mullard Space Science Laboratory, Holmbury St. Mary, Dorking, Surrey, RH5 6NT, UK}
\affiliation{LESIA, Observatoire de Paris, Universit\'{e} PSL, CNRS, Sorbonne Universit\'{e}, Univ. Paris Diderot, Sorbonne Paris Cit\'{e}, 5 place Jules Janssen, 92195 Meudon, France}
\affiliation{Konkoly Observatory, Research Centre for Astronomy and Earth Sciences, Hungarian Academy of Sciences, Konkoly Thege \'{u}t 15-17., H-1121, Budapest,
Hungary}

\author[0000-0002-3362-7040]{J. Martin Laming}
\affiliation{Space Science Division, Naval Research Laboratory, Code 7684, Washington, DC 20375, USA}

\author[0000-0001-7809-0067]{Gherardo Valori}
\affiliation{University College London, Mullard Space Science Laboratory, Holmbury St. Mary, Dorking, Surrey, RH5 6NT, UK}



\begin{abstract}

% Elemental abundances in the solar corona are different from those in the photosphere and the differences seem to be strongly modulated by the magnetic field and magnetic activity. Elemental abundances of easy-to-ionise low First Ionisation Potential (FIP) elements with FIP below 10 eV are enhanced in active region coronae compared to their photospheric abundances, while high FIP elements (FIP $\geq$ 10 eV) maintain their photospheric abundances, a phenomenon known as the FIP effect. Observed changes in elemental abundances take place on the time scales of years (solar cycle), days or hours (magnetic flux emergence and decay), or minutes (flare). 

We analyse the coronal elemental abundances during a small flare using Hinode/EIS observations. Compared to the pre-flare elemental abundances, we observed a strong increase in coronal abundance of \ion{Ca}{14} 193.84~\AA, an emission line with low first ionisation potential (FIP $<$ 10~eV), as quantified by the ratio Ca/Ar during the flare. This is in contrast to the unchanged abundance ratio observed using \ion{Si}{10} 258.38~\AA/\ion{S}{10} 264.23~\AA. We propose two different mechanisms to explain the different composition results. Firstly, the small flare-induced heating could have ionised S, but not the noble gas Ar, so that the flare-driven Alfv\'{e}n waves brought up Si, S and Ca in tandem via the ponderomotive force which acts on ions. Secondly, the location of the flare in strong magnetic fields between two sunspots may suggest fractionation occurred in the low chromosphere, where the background gas is neutral H. In this region, high-FIP S could behave more like a low-FIP than a high-FIP element. The physical interpretations proposed generate new insights into the evolution of plasma abundances in the solar atmosphere during flaring, and suggests that current models must be updated to reflect dynamic rather than just static scenarios.

% Although theories have been proposed to explain this effect, the underlying physics of the evolution of solar coronal composition is only starting to see its light recently. 
% , which has typically been identified as a high-FIP element
\end{abstract}

\keywords{Sun: abundances - Sun: corona - Sun: magnetic fields}


\section{Introduction} \label{intro}

Composition of plasma in the solar corona is a tracer of the flow of plasma and energy from the solar interior.  Different complex processes such as the propagation and absorption of waves, convection of hot plasma and reconnection and reconfiguration of magnetic fields can affect the flow and composition of plasma. This produces a clear and observable variation in the elemental abundances of coronal plasma across different regions of the solar atmosphere \citep[\eg][]{Brooks2015}.

In order to parameterise and study the coronal elemental abundances, we use the first ionisation potential (FIP) bias, defined as taking the ratio of an element's coronal to photospheric abundance with respect to H. The FIP bias varies from solar structure to solar structure, and is closely linked to the Sun's magnetic field on all scales. Observed changes in elemental abundances take place on time scales of years (solar cycle), days or hours (magnetic flux emergence and decay), or minutes (flare) \citep{Brooks2017,Baker2018,Warren2014Apr}. In a quintessential active region, low-FIP elements (FIP $<10$~eV) such as Ca, Mg and Si exhibit enhanced abundances, while high-FIP elements (FIP $\ge10$~eV) such as Ar, O and S retain their photospheric abundances. This elemental fractionation is known as the FIP effect. The FIP bias of the photosphere is typically measured to be 1, with higher FIP bias values of 3-4 found in the closed loops of a developed active region \citep[\cf][]{DelZanna2014May,Baker2013,Baker2015,Baker2018}. The FIP bias value of quiet-Sun \citep{Warren1999Dec,Baker2013,Ko2016Aug} and coronal hole regions \citep{Feldman1998Oct,Brooks2010Dec,Baker2013} has been explored extensively, and has also been used to link solar wind back to its source regions \citep{Gloeckler1989,Fu2017Feb}

In addition to the small magnetic perturbations that are common in a typical active region, rapid changes in magnetic connectivity can also contribute to a change in coronal composition. One of the events that causes rapid coronal composition changes is a solar flare. Solar flares are spectacular results of solar magnetic reconnection, characterised by the sudden release of energy and plasma. This rapid change of magnetic configuration triggers waves and energy that propagate into the chromosphere \citep{Fletcher:2008}, followed by the ablation of plasma into the corona \citep{Warren2014Apr}. This produces the variability in coronal emission intensities observed by spectrometers. The standard and static coronal composition picture relies on a selective mechanism, generating a coronal composition that is different from the photospheric one. On the other hand, ablation does not act on specific elements. \citet{Warren2014Apr} studied 21 M and X class flares using Solar Dynamic Observatory/EUV Variability Experiment (\textit{SDO}/EVE) showed that ablation acted on all elements and turned the Sun-as-a-star coronal elemental abundances temporarily closer to photospheric, in agreement with earlier results of \citet{Veck1981Oct, Feldman1990Nov, McKenzie1992Apr} and \citet{DelZanna2013Jul}. However, an explanation that only comprises of ablation falls short on addressing some observations that obtained an enhanced low-FIP elemental abundances \citep[e.g.][]{Doschek1985Nov,Sterling1993Feb,Bentley1997Jan,Fludra1999Aug,Phillips2010Feb,Phillips2012Mar,Dennis2015Apr,Sylwester2015May}, as well as the more recent observations that have found evidence of an inverse FIP (IFIP) effect during the decay phase of flares occurring in very complex active regions. In such cases, the IFIP signatures were observed with timescales of around 30 minutes \citep{Doschek2015Jul,Doschek2016Jun,Baker2019Apr, Baker2020May}.

So far, the only theoretical model that is able to explain both the FIP and IFIP effects is the ponderomotive force fractionation model, proposed by \citet{Laming2004Oct,Laming2009Apr,Laming2011Dec,Laming2015Sep}. In this model, FIP effect fractionation is caused by the reflection and refraction of Alfv\'{e}n waves in the upper chromosphere, especially in regions with high density gradients. The change in wave direction exerts a back reaction on the plasma ions, the ponderomotive force, and depending on the origin and nature of the waves this can vary in magnitude and sign. The direction of the ponderomotive force leads to ions that are guided in different directions, causing the FIP/IFIP effect.

In this paper, we present Hinode/Extreme-ultraviolet  Imaging  Spectrometer (\emph{Hinode}/EIS) observations taken during a solar flare which originated in a reconnecting X-shaped structure rooted in a complex active region. EIS catching the flare in action provides an insight into the rapid evolution of solar coronal composition. The observations are presented in Section~\ref{obs}, with results and discussion in Section~\ref{results} and Section~\ref{discussion}. Conclusions are then discussed in Section~\ref{conclusion}.

\section{Observations and Data Analysis}\label{obs}

\begin{figure*}[ht!]
    \centering
    \includegraphics[width=\textwidth]{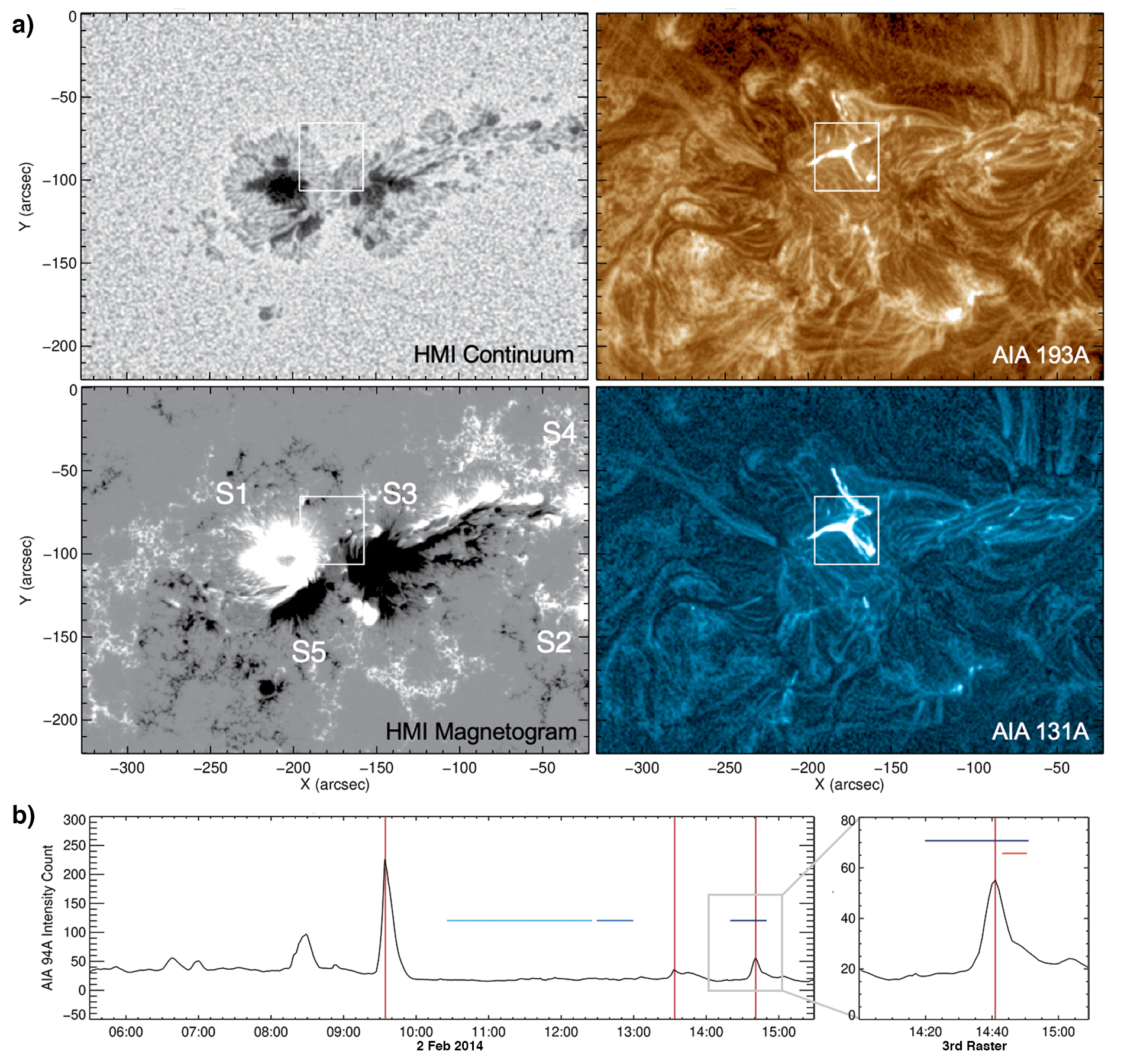}
    \caption{a) HMI continuum; HMI magnetogram; AIA 193~\AA; AIA 131~\AA\ image during the flare at 14:34~UT on 2~Feb~2014. The white box indicates the EIS field of view in Figure~\ref{fig:ca_ar_maps}. b) AIA 94~\AA\ light curve obtained using a 34\arcsec$\times$31\arcsec\ box surrounding the X-shaped structure in AR~11967. Blue horizontal lines indicate EIS rastering time and duration; Red vertical lines indicate the flare time; Orange horizontal line in the zoomed section of the light curve indicates the EIS scan time over the X-shaped structure. The third flare happened during the third EIS raster, and EIS observed the X-shaped structure during its decay phase, roughly 1 minute after the peak of the flare.}
    \label{fig:flare_AIA}
\end{figure*}

AR 11967 was an old and complex active region with a rich history. It was visible on the Southern hemisphere of the Sun from 31~January~2014 to 7~February~2014. The active region was in its second rotation on disk, and major flux emergence and reconnection could be observed during this rotation. Figure~\ref{fig:flare_AIA} shows the magnetic complexity of AR~11967 from the Solar Dynamics Observatory/Helioseismic and Magnetic Imager \citep[\emph{SDO}/HMI;][]{Hoeksema2014} line-of-sight magnetic field data. Its complex and active nature led to numerous eruptions, resulting in 28 M-class flares and 83 C-class flares. 

AR~11967 was particularly interesting as it hosted a very stable, long lived X-shaped structure \citep[previously identified and discussed by][]{Jiang2017,Kawabata2017,Liu2016,Xue2017,Yang2017}. This structure was observed from 2-5~Feb~2014, and was located between the sunspots S1 and S3 shown in the HMI magnetogram of Figure~\ref{fig:flare_AIA}.

\subsection{Coronal EUV and Magnetic Field Observations}
The active region could be identified and studied on its evolution across the disk using the full-disk Extreme-Ultraviolet (EUV) and line of sight magnetogram images from the Atmospheric Imaging Assembly \citep[AIA;][]{Lemen2012} and the HMI instruments on board the \textit{Solar Dynamics Observatory} \citep[SDO;][]{Pesnell:2012} spacecraft. Three different passbands from AIA were used in this analysis; 94~\AA, 193~\AA\ and 131~\AA. Images from the 193~\AA\ passband are especially useful for FIP bias analysis as they capture emission from elements at around the same temperatures as one of the FIP bias pairs, \ion{Si}{10} (258.38~\AA) and \ion{S}{10} (264.23~\AA), formed at $\sim$1.25-1.5~MK. In contrast, 131~\AA\ images are used as context images, as this passband benefits from clear flaring illumination, while 94~\AA\ images were used to create the averaged intensity light curve. The HMI line-of-sight magnetic field images were used to demonstrate the magnetic evolution and flux emergence of the active region. Both the AIA and HMI data were processed using the standard aia\_prep.pro routine within SolarSoftWare \citep{Freeland:1998}. This accounts for the differences in the plate scales, roll angles between AIA and HMI and correctly aligns the two instruments. Both 193~\AA\ and 131~\AA\ broadband images were also sharpened using the Multi-scale Gaussian Normalisation technique \citep[MGN;][]{Morgan2014Aug}.

\subsection{EUV Spectroscopic Observations}

The elemental abundance results discussed here were derived using spectroscopic observations from the Extreme-ultraviolet Imaging Spectrometer \citep[EIS;][]{Culhane2007Jun} onboard the \emph{Hinode} spacecraft \citep{Kosugi:2007}. Hinode/EIS observed AR~11967 on 2~Feb~2014, making three observations using two active region studies, study acronym HPW021\_VEL\_240x512v1 and Atlas\_30~(Study number 437 and 403 respectively). Table~\ref{table:study_details} shows the key details of the studies that were used to track the evolution of AR~11967. Study 437 has a large field of view of 240\arcsec\ $\times$\ 520\arcsec. It uses the slit scanning mode with the 1\arcsec\ slit and 1\arcsec\ scan step size, with a long exposure time of 60~s at each step for a total rastering time of 2~hours. In contrast, study 403 is a full CCD study, which has a smaller field of view of 120\arcsec\ $\times$\ 160\arcsec, using a 2\arcsec\ slit with a 2\arcsec\ step. At each pointing position, EIS exposed for 30~s to produce a raster with a duration of 30 minutes. Similar to the SDO data, the EIS data were preprocessed using the standard eis\_prep.pro routine available in SolarSoftWare; this accounts for dark current, CCD bias, cosmic rays, hot, warm, dusty pixels, the radiometric calibration, and orbital correction. The eis\_ccd\_offset.pro routine was then used to ensure spatial consistency between different EIS spectral windows. The two pairs of low-FIP/high-FIP elements presented here are close in wavelength, however the Fe lines used for Differential  Emission Measure~(DEM) span both EIS CCDs, so the \citet{DelZanna2013Jul_eis_cal} \textit{Hinode}/EIS calibration was used to calibrate the EIS data. In order to minimise the offset between \textit{Hinode}/EIS and \textit{SDO}, the intensity map of \ion{Fe}{12} 195.12~\AA\ was aligned by-eye using the AIA 193~\AA\ passband. Since EIS \ion{Fe}{12} 195.12~\AA\ and AIA 193~\AA\ sample plasma from similar temperatures, the same solar structures could be identified in both maps, making instrumental alignment easier and more accurate.

\subsection{Composition Maps}
In our FIP bias examination, two pairs of emission lines, \ion{Si}{10} 258.38~\AA/\ion{S}{10} 264.23~\AA\ and \ion{Ca}{14} 193.87~\AA/\ion{Ar}{14} 194.40~\AA, were used extensively to examine composition measurements at two different solar atmospheric temperatures. Both pairs of elements are comprised of one low-FIP element, Si~(Ca) and a high-FIP element, S~(Ar).

%  Both \ion{Si}{10}/\ion{S}{10} and \ion{Ca}{14}/\ion{Ar}{14} pairs are close in wavelength and contain no strong blends of other emission lines. 

\subsubsection{\ion{Si}{10}/\ion{S}{10} Composition Map}\label{composition_map_algorithm}
The \ion{Si}{10} (258.38~\AA, FIP = 8.15 eV) and \ion{S}{10} (264.23~\AA, FIP = 10.36 eV) line pair has a comparably lower formation temperature at around 1.25-1.5~MK. The emission from the consecutive ionisation stages of \ion{Fe}{8}--\ion{Fe}{16} were fitted with a single Gaussian function, with the exception of \ion{Fe}{11}, \ion{Fe}{12} and \ion{Fe}{13} fitted with multiple Gaussians to obtain their intensities. The effects of density variation were estimated using the fitted \ion{Fe}{13} 202.04~\AA/203.83~\AA\ line-pair ratio. These intensities were then used to derive the Differential Emission Measure~(DEM) so that we can account for temperature effects when modelling the \ion{S}{10} line intensity. Since the Fe lines lie on the short-wavelength detector and the Si and S lines lie on the long-wavelength detector, the emission measure was scaled to reproduce the intensity of \ion{Si}{10}~258.38~\AA.
The CHIANTI Atomic Database, Version 8.0 \citep{Dere1997Oct,DelZanna2015Oct} was used to obtain the contribution functions, applying the photospheric abundances of \citet{Grevesse2007Jun} for all the spectral lines, while assuming the \ion{Fe}{13} density calculated above. Lastly, both \ion{Si}{10}~258.38~\AA\ and \ion{S}{10}~264.23~\AA, which form our FIP bias ratio were fitted with a single Gaussian function. The Markov-Chain Monte Carlo (MCMC) algorithm from the PINTofALE software package was used to compute the emission measure distribution for the Fe lines \citep{Kashyap2000Jun}. The emission measure distribution was then convolved with the CHIANTI contribution functions and fitted to the observed spectral intensity of the low-FIP Fe emission lines. Since both Fe and Si are low-FIP elements, the best fit emission measure for each pixel will be enhanced by some factor due to the FIP effect. We can determine this factor by calculating the intensity of the \ion{S}{10}~264.23~\AA\ line; assuming S is a high-FIP element that is not enhanced. Finally, the Si/S FIP bias ratio presented in this paper is then the ratio of the predicted to observed intensity of the \ion{S}{10}~264.23~\AA\ line. In total, 17 spectral lines were used to infer our \ion{Si}{10} 258.38~\AA/\ion{S}{10} 264.23~\AA\ composition maps. Data pixels with a $\mathrm{\bar{\chi}^2} > 17$ were discarded during our analysis. Since EIS is formed by both a short-wavelength and long-wavelength detectors with an offset of 18.5 pixels in the y direction, part of our region of interest that lies at the top of the long-wavelength detector was not captured by the short wavelength detector. The lack of data for the \ion{Si}{10}~258.38~\AA/\ion{S}{10}~264.23~\AA\ composition calculation contributed to a horizontal strip of data with very high $\mathrm{\bar{\chi}^2}$ value, and this resulted in the patch of missing data shown in Figure~\ref{fig:ca_ar_maps}. The estimated uncertainty of the \ion{Si}{10}~258.38~\AA/\ion{S}{10} 264.23~\AA\ FIP bias ratio is 0.30, assuming a 20\% spectral line intensity error. This approach is discussed in detail by \citet{Brooks2015}, and is designed to remove the temperature and density effects for a robust calculation of the FIP bias in different solar features.

\subsubsection{\ion{Ca}{14}/\ion{Ar}{14} Composition Map}\label{ca_ar_comp_map}

\begin{figure*}[ht!]
    \centering
    \includegraphics[width=0.98\textwidth]{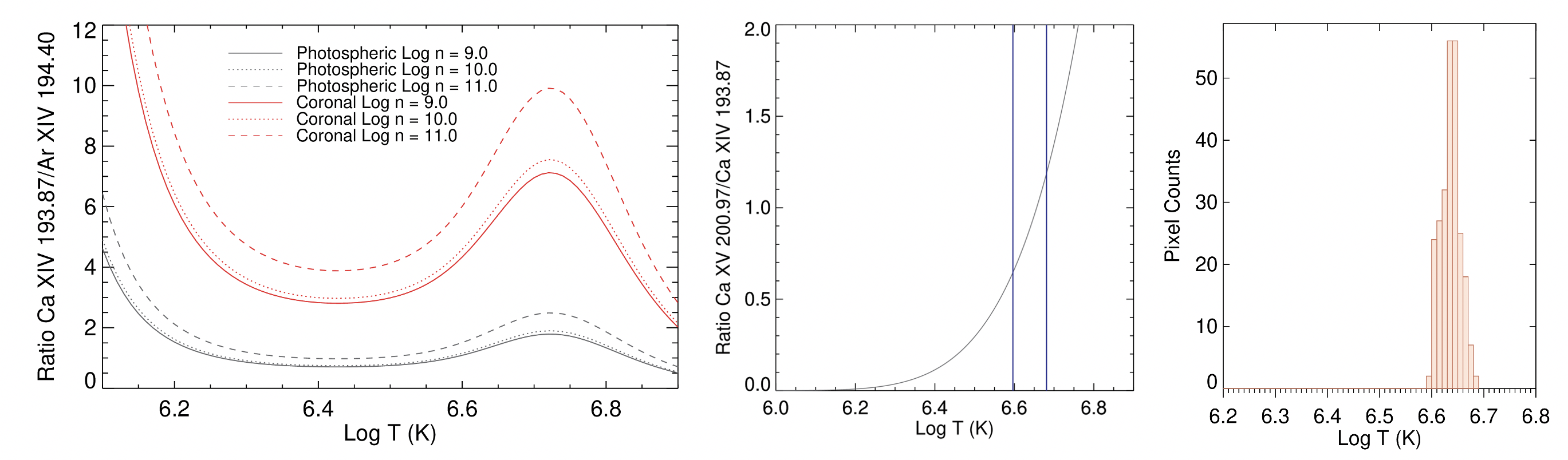}
    \caption{Left to right: a) Intensity ratio of \ion{Ca}{14}~193.87~\AA\ and \ion{Ar}{14}~194.40~\AA\ contribution function with respect to electron temperature for electron density of $\mathrm{log_{10}N = 9.0, 10.0, 11.0}$; b) Intensity ratio of \ion{Ca}{15}~200.97~\AA\ and \ion{Ca}{14}~193.87~\AA\ contribution function with respect to electron temperature; c) Temperature histogram around the flaring region.}
    \label{fig:gofnt}
\end{figure*}

The \ion{Ca}{14} (193.87~\AA, FIP = 6.11 eV) and \ion{Ar}{14} (194.40~\AA, FIP = 15.76 eV) emission lines are formed at a comparably higher ionisation equilibrium temperature of around 3.5~MK (log$_{\mathrm{10}}\mathrm{T}$ = 6.55) \citep{Feldman2009Mar}. Both lines are relatively strong and present similar emissivity temperature dependence. Three Gaussian functions were fitted to both the \ion{Ca}{14}~193.87~\AA\ and \ion{Ar}{14}~194.40~\AA\ lines, as \ion{Ca}{14}~193.87~\AA\ is sandwiched between two other lines, and sometimes \ion{Ar}{14} is blended with two very faint lines along its blue wing \citep{Brown2008}. In this paper, we follow the analysis done by \citet{Doschek2015,Doschek2016Jun, Doschek2017Jul,Baker2019Apr, Baker2020May} in producing the composition maps. In particular, \citet{Doschek2017Jul} describes the assumptions and issues that come with this composition diagnostic. We used $\mathrm{log_{10}}$ abundance values relative to $\mathrm{log_{10}H}$ as; coronal-Ca = 6.93 \citep{Feldman1992}, photospheric-Ca = 6.33 \citep{Lodders2009} and coronal and photospheric-Ar = 6.50 \citep{Lodders2009}. Since the spectral intensity of Ar at the photosphere could not be directly measured, its abundances value (photospheric-Ar) is determined indirectly through spectra observed in solar wind, solar flares or solar energetic particles. As a result, historically, there has been a relatively large fluctuation of the photospheric-Ar abundances. Figure~\ref{fig:gofnt} shows a plot of the ratio of \ion{Ca}{14}~193.87~\AA\ to \ion{Ar}{14}~194.40~\AA\ contribution functions at various densities using the abundances above. This yields a typical coronal active region FIP bias of 4 and has the advantage of being quick and simple. However, taking ratio between the low FIP \ion{Ca}{14}~193.87~\AA\ to the high FIP \ion{Ar}{14}~194.40~\AA\ does not account the effects of temperature and density. To alleviate the concerns over the ratio value being sensitive to these effects, temperatures were calculated using ratio between \ion{Ca}{15}~200.97~\AA\ and \ion{Ca}{14}~193.87~\AA\  around the region of interest; its temperature histogram is also shown in Figure~\ref{fig:gofnt}. The Ca/Ar composition maps observed before and during the flare are similar in temperature at $\sim\mathrm{log_{10}T=6.65K}$. The range of electron temperatures during the flare was also narrow, and ranges between $\mathrm{log_{10}6.59-log_{10}6.68}$. When these two pieces of evidence are put side by side with the significant intensity ratio changes observed across the Ca/Ar maps, the temperature and density can be seen to have a minimal effect for our analysis. The contribution functions that are convolved with the DEM at the flaring region is shown in the \nameref{Appendix}.

\subsection{Region Definition}

In Figure \ref{fig:ca_ar_maps}, we can observe a significant increase in the Ca/Ar intensity ratio value in our third raster (panel~f) corresponding to the stable structure, which clearly describes its X shaped morphology. Since this X-shaped structure is extremely stable throughout the period studied here on 2~Feb~2014, we defined the flaring region using pixels with an intensity ratio value $>$3 in this raster. This intensity ratio value represents an enhanced FIP bias. The pixels corresponding to this region of interest were then differentially rotated to the times of the two previous rasters on 2 Feb 2014, at 10:25~UT and 12:29~UT. The white border in each panel of Figure~\ref{fig:ca_ar_maps} shows the extent of this region of interest differentially rotated to the corresponding raster time. In each case there is good agreement, indicating the stability of the X-shaped structure and the robust nature of the method used to define the region of interest.

\begin{centering}
\begin{table}[]
\resizebox{\columnwidth}{!}{\begin{tabular}{@{}lll@{}}
\multicolumn{3}{c}{} \\ 
\toprule
Study Number & \multicolumn{2}{l}{437} \\
Study Acronym & \multicolumn{2}{l}{HPW021\_VEL\_240x512v1} \\
\multicolumn{1}{c}{} & \multicolumn{2}{l}{\ion{Fe}{8} 185.213~\AA, \ion{Fe}{8} 186.601~\AA} \\
\multicolumn{1}{c}{} & \multicolumn{2}{l}{\ion{Fe}{9} 188.497\AA, \ion{Fe}{9} 197.862~\AA} \\
\multicolumn{1}{c}{} & \multicolumn{2}{l}{\ion{Fe}{10} 184.536~\AA, \ion{Fe}{11} 188.216~\AA} \\
                     & \multicolumn{2}{l}{\ion{Fe}{12} 192.394~\AA, \ion{Fe}{12} 195.119~\AA} \\
Emission Lines       & \multicolumn{2}{l}{\ion{Fe}{13} 202.044~\AA, \ion{Fe}{13} 203.826~\AA}  \\
                     & \multicolumn{2}{l}{\ion{Fe}{14} 264.787~\AA, \ion{Fe}{14} 270.519~\AA} \\
                     & \multicolumn{2}{l}{\ion{Fe}{15} 284.16~\AA, \ion{Fe}{16} 262.984~\AA} \\
                     & \multicolumn{2}{l}{\ion{Fe}{17} 254.870~\AA} \\
                     & \multicolumn{2}{l}{\ion{Si}{10} 258.38~\AA, \ion{S}{10} 264.23~\AA} \\
                     & \multicolumn{2}{l}{\ion{Ca}{14} 193.87~\AA, \ion{Ar}{14} 194.40~\AA} \\
Field of View        & \multicolumn{2}{l}{240\arcsec\ $\times$ 512\arcsec} \\
Rastering            & \multicolumn{2}{l}{1\arcsec\ slit, 120 positions, 2\arcsec\ coarse steps} \\
Exposure Time        & \multicolumn{2}{l}{60~s} \\
Total Raster Time    & \multicolumn{2}{l}{2 hours} \\
Reference Spectral Window   & \multicolumn{2}{l}{\ion{Fe}{12} 195.12~\AA} \\  
\midrule
Study Number    & 403 \\
Study Acronym &  Atlas\_30 \\
Emission Lines & \multicolumn{2}{l}{\ion{Si}{10} 258.38~\AA, \ion{S}{10} 264.23~\AA} \\
               & \multicolumn{2}{l}{Ca\ion{}{14} 193.87~\AA, \ion{Ar}{14} 194.40~\AA} \\
Field of View     & 120\arcsec\ $\times$ 160\arcsec \\
Rastering         & 2\arcsec\ slit, 60 positions, 2\arcsec\ steps \\
Exposure Time     & 30~s \\
Total Raster Time & 30 minutes \\
Reference Spectral Window & \ion{Fe}{12} 195.12~\AA \\
\bottomrule

\end{tabular}}
\caption{\textit{Hinode}/EIS study details used in this study.}
\label{table:study_details}

\end{table}
\end{centering}

\section{Results}\label{results}

\begin{figure*}[h!]
    \centering
    \includegraphics[width=0.675\textwidth]{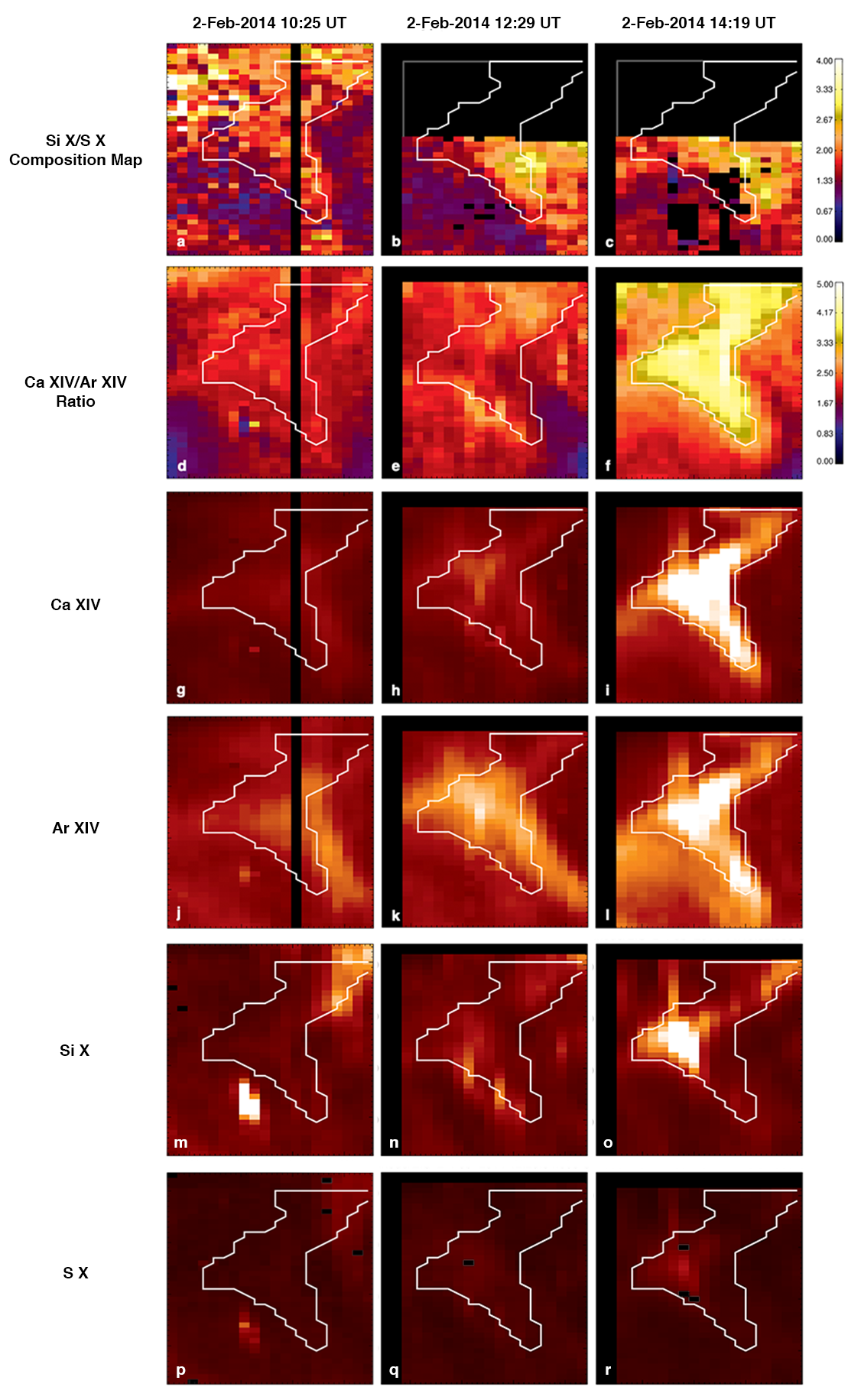}
    \caption{Top to bottom: Hinode/EIS \ion{Si}{10}~258.38~\AA/\ion{S}{10}~264.23~\AA\ composition map, \ion{Ca}{14}~193.87~\AA/\ion{Ar}{14}~194.40~\AA\ intensity ratio maps, \ion{Ca}{14}~193.87~\AA\ intensity maps, \ion{Ar}{14}~194.40~\AA\ intensity maps, \ion{Si}{10}~258.38~\AA\ intensity maps  and \ion{S}{10}~264.23~\AA\ intensity maps. Contour indicates the flaring region used for analysis. Left to right: observations from three different EIS spectral scans on 2~Feb~2014, starting at 10:25~UT, 12:29~UT and 14:19~UT, respectively. In the composition and intensity ratio maps, blue represents a photospheric-like (unenhanced composition) FIP bias of 1; red represents a quiet sun composition and yellow shows a high FIP bias composition. A clear X shape of the structure could be observed in the 14:19~UT raster. The algorithm to calculate the \ion{Si}{10}~258.38~\AA/\ion{S}{10}~264.23~\AA\ 
    composition map minimises temperature and density effects by utilising both the short-wavelength and long-wavelength detectors of \textit{Hinode}/EIS. Offset between the two detectors contributed to the patch of dark pixels in the composition maps; The \ion{Ca}{14}~193.87~\AA/\ion{Ar}{14}~194.40~\AA\ ratio maps are simply intensity ratio of the two spectral lines. A more detailed description can be found in Section \ref{composition_map_algorithm} and \ref{ca_ar_comp_map}.}
    \label{fig:ca_ar_maps}
\end{figure*}

Although the active region was observed by EIS between 1 and 5~Feb~2014, and the X-shaped structure was apparent between 2 and 4~Feb, here we focus on its evolution between 9:30~UT and 14:50~UT on 2~Feb. The bottom panel of Figure~\ref{fig:flare_AIA} shows the averaged intensity in the 94~\AA\ passband in the field of view shown by the white box in Figure~\ref{fig:flare_AIA}a. Multiple brightenings can be identified, with the 3 blue horizontal lines indicating the periods of the EIS rasters. The first flare was the strongest, starting at 09:26~UT, $\sim$1 hour before the start time of our first raster. It peaked at 09:35~UT, and decayed to background intensity levels by $\sim$09:51~UT. The second flare was a very minor flare as shown by the small peak in Figure~\ref{fig:ca_ar_maps}. It started at 13:22~UT, after our first two rasters and around 1 hour before the start time of our third raster. It peaked at 13:29~UT and decayed into background intensity levels by $\sim$13:53~UT. Finally, the third flare took place during the third EIS raster described here. The flare started at 14:27~UT, approximately 9 minutes into the raster, peaked at 14:37~UT and decayed to background intensity level by $\sim$14:56~UT. Neither flare 2 or 3 could be identified in the GOES X-ray flux, indicating that they were very minor flares.

\subsection{Magnetic Field Evolution}
\begin{figure*}[ht!]
    \centering
    \includegraphics[width=\textwidth]{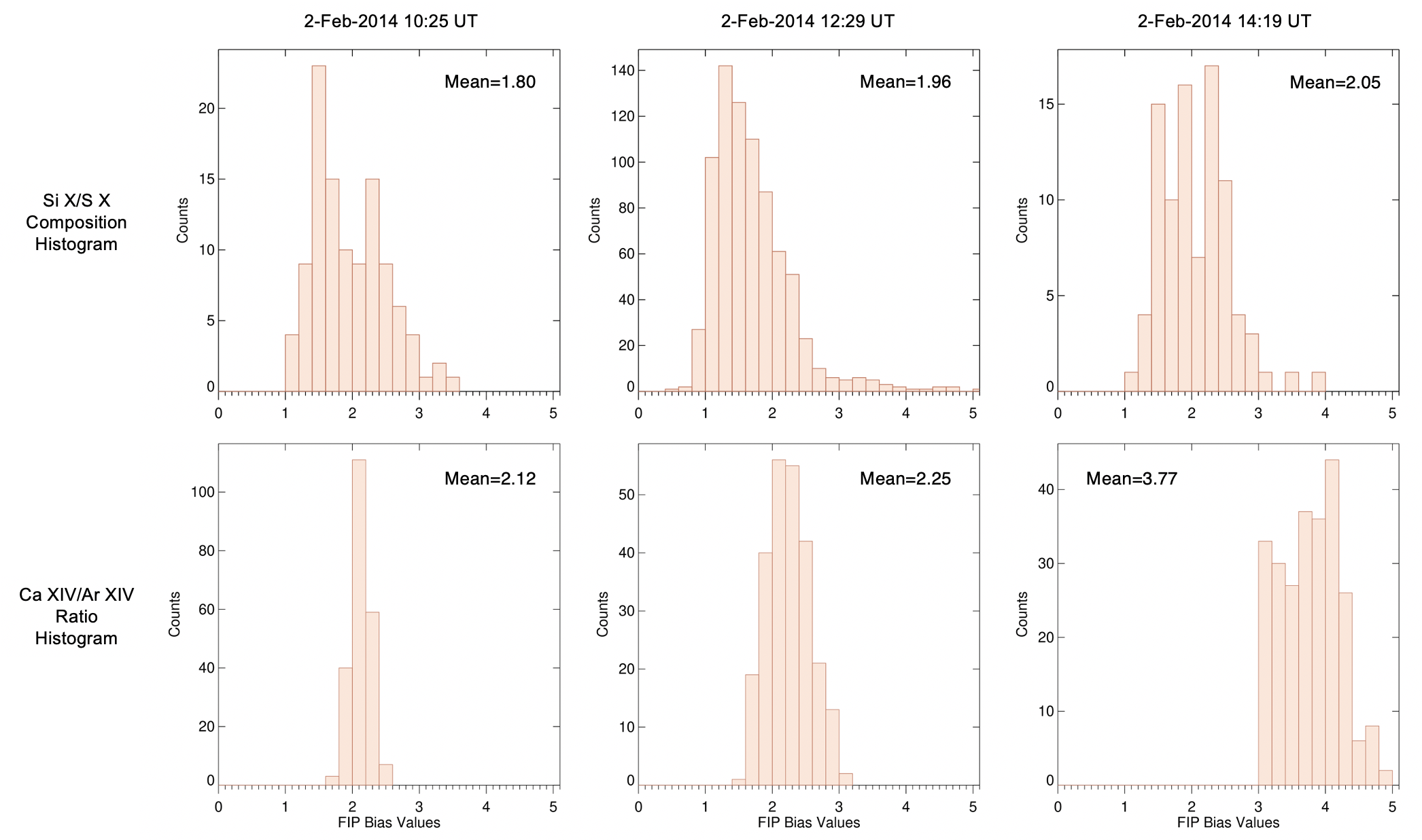}
    \caption{Top to bottom: Histogram of a) \ion{Si}{10}~258.38~\AA/\ion{S}{10}~265.23~\AA\ FIP bias values; b) \ion{Ca}{14}~193.87~\AA/\ion{Ar}{14}~194.40\AA\ intensity ratio values, at the region of interest highlighted with white contour in Figure~\ref{fig:ca_ar_maps}.}
    \label{fig:CaAr_FIP_hist}
\end{figure*}
AR 11967 was a highly complex AR built up by at least three major bipoles in various stages of their evolution. Figure~\ref{fig:flare_AIA} shows both S1 and S2, which were leading (positive) polarity spots produced by two flux emergence episodes before the AR rotated onto the disk. We can see that the negative S5 is located tightly below S1, but in fact S1 and S2's following (negative) polarities had already dispersed by this time. Despite this, S1 was still a large spot exceeding 100 MSH according to the Debrecen Photoheliographic Data. A major flux emergence occurred in the AR during the EIS observation period, with its principal spots being S4 (positive) and S3 (negative) in Figure~\ref{fig:flare_AIA}. The parallel strands of opposite polarities in between S4 and S3 indicate that this emerging bipole is strongly sheared. The magnetic polarity pattern \citep[magnetic tongues, \cf][]{Luoni:2011} is indicative of negative, left-handed twist. As S4 and S3 are separating in the course of flux emergence, the negative polarity S3 is moving towards the positive polarity S1. As the flare observed with EIS took place in between spots S3 and S1 (\cf\ Figure~\ref{fig:flare_AIA}), a forcing by their persistent approach was creating strong field gradients in the proximity of the X-shaped structure~\citep{Jiang2017, Kawabata2017}.

\subsection{Enhanced Low FIP Elemental Abundances of the Third Flare}

Figure~\ref{fig:ca_ar_maps} shows the \ion{Si}{10}/\ion{S}{10} composition map and the \ion{Ca}{14}/\ion{Ar}{14} intensity ratio map from the three EIS rasters made on 2~Feb~2014. EIS made two raster observations after the first flare, with the second flare occurring before, and the third flare during, the third EIS raster. It can be seen in the first row of Figure~\ref{fig:ca_ar_maps} that, across the three observations, in a temperature range $\sim$1.25-1.5~MK, the \ion{Si}{10}/\ion{S}{10} composition map does not show drastic changes, and the map is generally red/dark orange in colour, comparable to a composition map of a quiet sun region or a small active region.

However, at a higher temperature (corresponding to the \ion{Ca}{14}/\ion{Ar}{14} ratio maps), a significant enhancement in FIP bias can be observed during the third flare. For the first two rasters, the region of the solar flare shows up as red in the intensity ratio map, indicating an ordinary quiet sun FIP bias value of ~2, similar to the FIP bias value calculated using the \ion{Si}{10}/\ion{S}{10} composition map. However, a clear change in coronal abundances can be seen in the third raster, where EIS observes the third flare (as a yellow region, indicating enhanced FIP) which peaked at 14:29~UT. The shape of the enhanced intensity ratio patch fully traces the topology of the X-shaped flare structure. 

These changes in the maps are reflected in the data histograms in Figure~\ref{fig:CaAr_FIP_hist} taken around the flaring region. The \ion{Si}{10}/\ion{S}{10} histogram shows little change between the three observations. On the other hand, the \ion{Ca}{14}/\ion{Ar}{14} ratio value histogram shows a drastic enhancement in the mean FIP bias value, which changes from a quiet sun value of $\sim$2 to 3.77, which is well beyond the estimated error. The FIP bias value therefore exhibits a significant difference between two different temperatures.

% In order to further investigate the intensity ratio values, every high ratio value pixels >3 in the third EIS raster were highlighted and rotated to their previous raster to obtain the intensity ratio values of the QSL prior to the flare. Figure \ref{fig:CaAr_FIP_hist} contains the histogram of the FIP bias value at the highlighted pixels for the three rasters. 

\section{Discussion and Interpretation}\label{discussion}
In this study, we have analysed the instantaneous coronal plasma evolution of a small flare in a reconnecting X-shaped structure located in the highly active AR~11967 observed by \emph{Hinode}/EIS on 2~Feb~2014. This opened up a possibility of obtaining composition during a solar flare. Over the course of the three observations, a low FIP bias value of $\sim$2 can be seen across all of the Si/S composition maps. Although the higher temperature Ca/Ar composition map shows comparable behaviour for the first two rasters, it shows a significantly enhanced ratio value for the third raster in both the composition map and the ratio value histogram in Figure~\ref{fig:CaAr_FIP_hist}, indicating a strong FIP effect during the flare. This inconsistent behaviour between composition maps formed using two line pairs is very interesting as it is different from the photospheric composition in flares found by \citet{Warren2014Apr,Baker2019Apr}, which is typically interpreted as an enhancement of photospheric material in the corona due to chromospheric ablation as part of the flare process.

% Moreover, the increased ratio value patch observed in the \ion{Ca}{14}/\ion{Ar}{14} map traces the topology of the QSL structure where reconnection happened, hinting that the stable, yet complex magnetic structure contributed to the observed enhanced FIP effect.

\subsection{Physical Interpretation}

We suggest two possible scenarios that could help to explain the evolution of the composition, in addition to the ablation picture we expect after flares. First, the fact that S as a high-FIP element has a relatively low FIP value which is close to 10~eV. Secondly, the potential shift of elemental fractionation height due to the strong magnetic fields observed here.

% \subsection{Flare-driven ablation of ionised chromospheric plasma}
The model to explain the observed fractionation of elemental abundances in the solar atmosphere is a quasi-static model of long-lasting structures such as coronal loops, active regions and solar wind outflow regions \citep[cf.][]{Laming2004Oct,Laming2019Jul}. Under these static scenarios, in closed loops, the fractionation occurs at the top of the chromosphere. \citet{Warren2014Apr} studied the evolution of plasma composition in large flares which involves a dynamic rapid evolution of plasma parameters (highly time dependent), and suggested that in some dramatic time-dependent events like large M and X class flares, the observed elemental abundance variation was due to plasma ablated from deep in the chromosphere, below where fractionation occurs. However, this is not the case here, where an enhanced abundance of Ca, or in other studies, where an enhancement of other various low-FIP elements \citep[e.g.,][]{Fludra1999Aug,Phillips2010Feb, Phillips2012Mar,Dennis2015Apr} was observed. Therefore, we have to look for other mechanisms to find an explanation.

% , which produced a heating effect much smaller than that observed by \citet{Warren2014Apr}, this mechanism could explain the observed S and lack of observed Ar. S is fractionated in the upper chromosphere, whereas Ar is fractionated in the lower chromosphere \citep{Laming2019Jul}, with the result that the low heating effect of the weak flare could have ablated S, but not reached deep enough in the chromosphere to ablate Ar. However, although this scenario may have been a contributing factor to the anomalous observations described here, a mechanism with only elemental ablation falls short of explaining the enhanced abundances of Ca, and various low-FIP elements in other studies 

\subsubsection{Partial Ionisation of Different Elements}
Firstly, the different composition evolution in the Si/S and Ca/Ar line pairs could be due to the actual FIP values of these elements. We have adopted the convention of categorising Si and Ca as low FIP elements and S and Ar as high FIP elements. While S is typically categorised as a high FIP element, it has a relatively low FIP value of 10.36~eV, whereas Si, typically categorised as a low FIP element, has a comparable FIP value of 8.15~eV. The difference between their FIP values is therefore merely 2.21~eV. In contrast, as Ar is a noble gas, it has a much higher FIP of 15.76~eV, with the result that the difference in FIP between it and Ca (FIP = 6.11~eV) is 9.65~eV. 

During a flare, magnetic reconnection produces Alfv\'{e}n waves that travel from the corona to the chromosphere \citep[e.g.][]{Fletcher:2008}. The refraction of such waves at the top of the chromosphere can then create an upward ponderomotive force that acts only on ions \citep[cf.][]{Laming2009Apr}. However, the same reconnection process also accelerates and heats particles, which at their impact onto the chromosphere lead to heating and consequent ablation of plasma. For fractionation to occur, the Alfv\'{e}n speed must be greater than the electron thermal speed, so that following coronal energy release, Alfv\'{e}n waves get to the chromosphere first to cause fractionation before the heat conduction flux arrives to cause the evaporation. If this is violated, unfractionated plasma will be evaporated into the flare loops. However, if the two speeds were about equal, fractionation with extra ionisation of S would result. For the flare studied here, which was very small, the heating produced could have been sufficient to ionise S with its relatively low FIP, with the ponderomotive force bringing up both Si and S in tandem, thus maintaining the quiet-sun FIP bias values that can be seen across the Si/S composition maps and their respective data histograms. On the contrary, due to the relatively high FIP of Ar, the small heating associated with the observed small flare was insufficient to ionise Ar. Instead, the more abundant low FIP Ca ions were brought up by the ponderomotive force, thus generating the fractionation observed in the third Ca/Ar intensity ratio map.

Within the above interpretation, this observation could open up a unique way to probe into the fractionation height in the solar atmosphere by different solar phenomena. With the partial ionisation of S but not Ar, one can estimate the upper chromospheric temperature at the time of the flare. Using \citet{Laming2019Jul}, the chromospheric model by \citet{Avrett2008Mar} and collisional ionisation equalibrium calculation from \citet{Mazzotta1998Dec}, the chromospheric temperature needs to be
\begin{equation}
    2.5\times\mathrm{10^4K< T < 3.0\times10^4K}.
\end{equation}
The true temperature is likely to be lower, because calculations from \citet{Mazzotta1998Dec} neglect photo-ionisation and density effects. Nonetheless, this postulate provides an additional insight into the work done by \citet{Warren2014Apr} and \citet{Dennis2015Apr}, who studied the composition evolution in large flares. For larger flares, more energy is released, which should produce a correspondingly large increase in temperature, thus ionising Ar. This would then produce sufficient ionised Ar to experience the ponderomotive force, resulting in the photospheric-like coronal composition more commonly observed during flares.

\subsubsection{Fractionation in the Low Chromosphere}

Another plausible scenario comes from the fact that the flare studied here took place between the strong magnetic fields of two large sunspots. This has the effect of lowering the plasma $\mathrm{\beta=1}$ height and consequently the fractionation height of the different elements. As identified in \citet{Fletcher:2008}, a solar flare which triggers reconfiguration of the magnetic field generates large-scale Alfv\'{e}n waves which propagate from the flaring site to the lower chromosphere, where the background gas is neutral H. In this region, high FIP S behaves like a low FIP element \citep{Laming2019Jul}; behaviour consistent with the statistically insignificant changes of Si/S FIP bias observed here. This requires a relaxing of the quasi-static nature of FIP models, to allow fractionation to occur throughout the chromosphere and not just at the top, which follows if the waves are in resonance with the loop. As with the scenario involving partial ionisation described above, this would result in fractionation of Ca and both Si and S, producing the constant Si/S and an increase in Ca/Ar observed here. These observations are consistent with the ponderomotive force interpretation for variation in FIP bias.

\subsection{AIA Wavelet Analysis}
\label{ss:wavelet}

\begin{figure*}[ht!]
    \centering
    \includegraphics[width=0.98\textwidth]{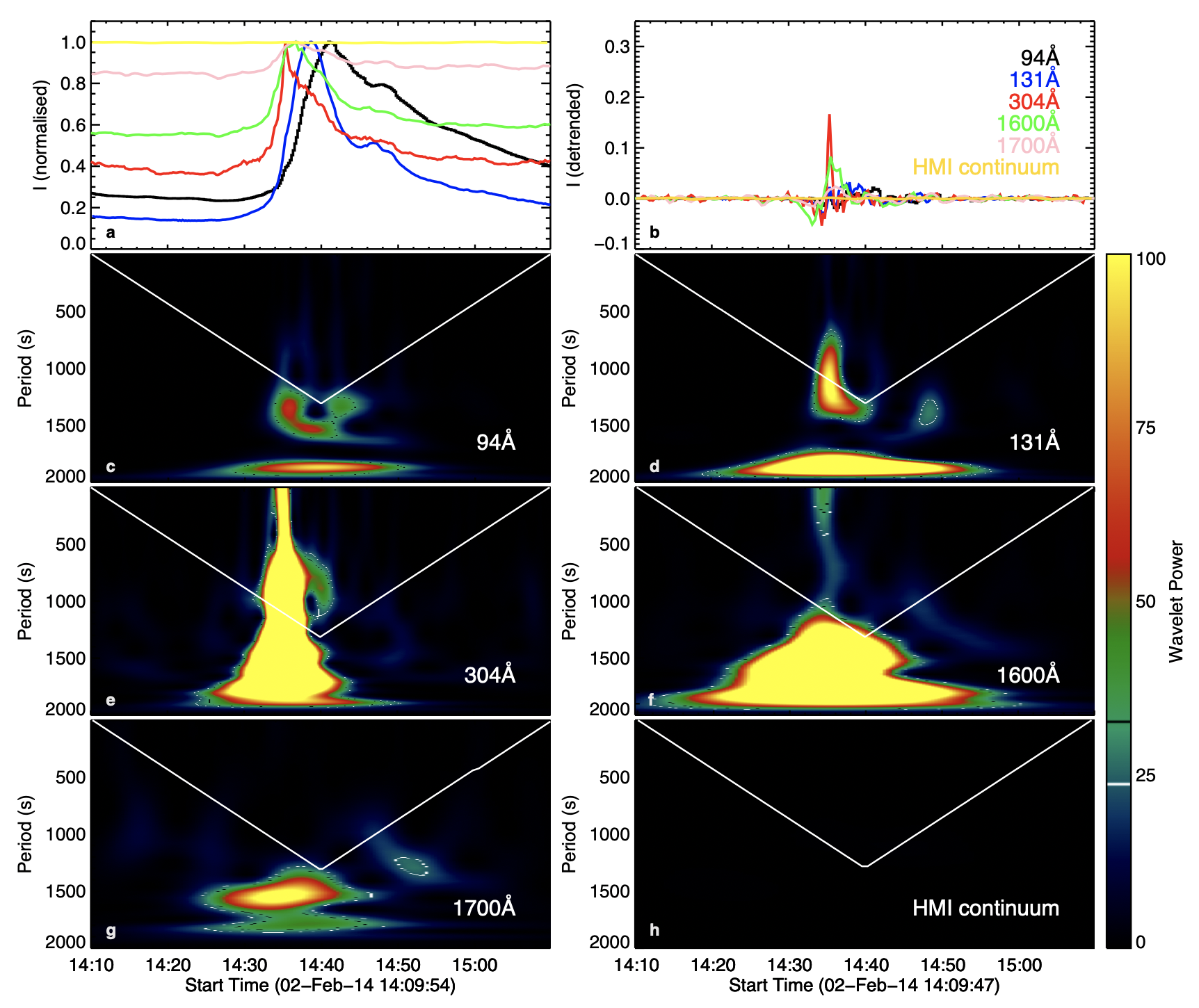}
    \caption{Wavelet analysis of AIA 94~\AA, 131~\AA, 304~\AA, 1600~\AA, 1700~\AA\ and HMI continuum during the third flare at 14:37~UT.}
    \label{fig:wavelet}
\end{figure*}

Although the two physical interpretations both seem plausible, they both rely on flare-driven downward travelling Alfv\'{e}n waves which induce the ponderomotive force acting upwards. To try and identify this essential signature of Alfv\'{e}n waves propagating to the chromosphere, the AIA data in different passbands were analysed using a wavelet technique similar to that previously used by \eg\ \citet{Milligan:2017,Hayes2016Aug}. The intensity in the HMI continuum, 1600~\AA, 1700~\AA, 304~\AA, 131~\AA\ and 94~\AA\ passbands was averaged within the region indicated by the white box shown in Figure~\ref{fig:flare_AIA}. The top left of Figure~\ref{fig:wavelet} shows the normalised wavelet analysis of the wavelengths, with the top right panel showing the detrended data produced by subtracting a smoothed version of the data. This was then analysed using the wavelet technique of \citet{Torrence:1998}. It can be seen in panels c--h of Figure~\ref{fig:wavelet} that each of the different passbands (with the exception of the HMI continuum) exhibits a variation in the signal associated with the small-scale brightening discussed here. However, the signals are in the form of a single peak instead of an oscillation. Moreover, in terms of the signal strength, apart from 304~\AA, no other passbands show a significant intensity. It is also notable that most of the observed signal for each passband is below the white triangular lines, indicating that they are not statistically significant and outside the zone of influence. 

This suggests that the Alfv\'{e}n waves produced by the flare in the corona proposed by \citet{Fletcher:2008} could not be observed in the lower solar atmosphere using the data available here. The lack of a statistically sound signal that could support wave propagation is disappointing, but this result could still inform several investigation directions for future work. Firstly, the instrument, AIA, used in the analysis has a relatively longer cadence of 12~s, which places a limit on the frequency of the observable signal. Using an instrument with a higher cadence, such as the \textit{Geostationary Operational Environmental Satellite's} X-ray sensor (GOES/XRS) could show the finer scale oscillations, albeit without the spatial resolution required here. A back of the envelope calculation gives insight into the wavelet period we can look into in the future, using a simple calculation from the coronal loop length and $\mathrm{\mathbf{v_A}}=\textbf{B}/\sqrt{\mu_0\rho}$, where $\mathrm{\mathbf{v_A}}$ is the Alfv\'{e}n speed. In our case, taking from the observations and assuming a semi-circle coronal loop, the loop length can be estimated to be $\sim$40\arcsec; and the x-shaped structure is rooted in an intense magnetic field with a line of sight field strength of $\sim$2,500 Gauss. Assuming a loop density of $\sim3\times10^{-7}~\mathrm{kgm^{-3}}$ \citep{Kohutova2017Jun}, this gives the line of sight Alfv\'{e}n speed, $\mathrm{v_{Az}}$, to be $\sim$400~$\mathrm{kms^{-1}}$, and a resonant wave period of 140 seconds. This value can then be compared with future wavelet analysis and help pinpoint the Alfv\'{e}n waves.

Secondly, AIA is a broadband imager, which captures spectral lines across a wide range of ions. Although a strong signal can be found in the 304~\AA\ wavelet analysis, which seems to suggest variations indeed exist in the chromosphere, the contributing spectral lines could not be precisely identified. The use of co-temporal narrow band spectroscopic data could provide more insight into the wave propagation depth, and perhaps open up the possibility of correlating the oscillation strength with the the observed FIP bias.

% A slight offset is apparent in the time between peak intensities measured for each passband, with the signal peaking first in the 304~\AA\ passband, followed by 1600~\AA, 1700~\AA, 131~\AA\ and finally 94~\AA. However, it is worth noting that the 94~\AA, 1700~\AA\ and HMI continuum passbands exhibit no statistically significant signal (indicated by signal within the white diagonal lines in each plot). This suggests that waves produced by the flare in the corona propagated through the transition region towards the chromosphere and upper photosphere, consistent with the interpretation of \citet{Fletcher:2008}. However, the lack of signal in the 1700~\AA\ and HMI continuum suggests that the waves did not reach the mid-low photosphere, with the lack of signal in the 94~\AA\ passband indicating that the flare was not particularly strong. Following the production of these waves, the delayed peak in 131~\AA\ intensity suggests that the heating followed much later. This is consistent with the ponderomotive force model proposed by \citet{Laming2004Oct}, with the waves driving the ponderomotive force originating from the reconnection site and propagating towards the footpoints of the magnetic structure described here.

% \begin{figure*}[ht!]
%     \centering
%     \includegraphics[width=\textwidth]{SiS_FIP_hist.png}
%     \caption{Histogram of \ion{Si}{10}/\ion{S}{10} FIP bias values at the QSL.}
%     \label{fig:SiS_FIP_hist}
% \end{figure*}

\section{Conclusions}\label{conclusion}
In this paper we have presented observations of the highly complex active region, AR~11967, which have provided a unique opportunity to study the evolution of composition in a small flare that occurred between two large sunspots. On 2~Feb~2014, EIS observed a small flare within an X-shaped structure in the active region. The results show very different composition evolution at two different line pairs of different formation temperatures, with an unchanged composition obtained in the lower temperature \ion{Si}{10}/\ion{S}{10} composition map in the flaring region, while a significant FIP bias value increase was observed in the higher temperature \ion{Ca}{14}/\ion{Ar}{14} composition ratio. 

We propose two possible physical interpretations which could explain or contribute to the strange composition evolution. Firstly, in the case of partial ionisation, due to the relatively low first ionisation potential of S, both Si and S were ionised by the small third flare with relatively low chromospheric heating, leaving the \ion{Si}{10}/\ion{S}{10} FIP bias value unchanged. While the lower FIP of S meant that it could be ionised by the small flare observed here, the much higher FIP value of Ar meant that it could not be easily ionised. The resulting ponderomotive force produced by the waves originating from magnetic reconnection leading to the flare thus only brought up the ionised Ca, Si and S in tandem. A similar yet non-mutually exclusive mechanism could be present in our second interpretation, fractionation at the lower chromosphere. Our flare occurred between two sunspots, rooted in a region of strong magnetic field. This has the effect of lowering the regions of fractionation of the different elements. Alfv\'{e}n waves generated by the flare travelled to the lower chromosphere, where the background gas is neutral H. Under this condition, S behaves like a low-FIP element. Therefore both Si and S fractionate in this flare and the Si/S ratio does not change, while the high-FIP Ar does not change its behaviour. This creates a significant discrepancy that can be observed between the two sets of composition maps. Both interpretations of the evolution of composition during a small flare that we present here are consistent with the ponderomotive force interpretation for variation in FIP bias.

Since these two interpretations involve Alfv\'{e}n waves that travel to the chromosphere, a wavelet analysis approach applied to the co-temporal AIA data was used to search for wave signatures. However, waves in the lower chromosphere during the reconnection were not detected with this method. Nonetheless, this provides the further investigation direction at correlating elemental fractionation with wave oscillations. Firstly, the relatively long cadence of AIA (12~s) limits the observation of very high frequency signals. Secondly, AIA/\textit{SDO} is a broad band imager. Signals obtained from different passbands merely give a very arbitrary idea of of the solar altitude. The use of co-temporal narrow band spectrometer will give more insight of the wave propagation depth, and perhaps open up the possibility of correlating the oscillation strength with the observed FIP bias. Some of the work to combine observations between different layers of the solar atmosphere has been done in \citet{Baker2021Jan}, using the Interferometric BIdimensional Spectrometer (IBIS), EIS and magnetic field modelling. Ground-based instruments like IBIS and Daniel K. Inouye Solar Telescope (DKIST) will be extremely valuable in the future, by directly observing where the wave refraction and reflections are proposed to happen. The upcoming Solar-C EUVST and its wide range of temperature coverage can also contribute massively by observing different layers of our Sun's atmosphere simultaneously. If the existence of Alfv\'{e}n waves can be confirmed, this observation could be the another step to understand the physical mechanism behind composition evolution during flares, and current theories need to factor the time evolution of composition into their modelling.

\section{Appendix}\label{Appendix}

\begin{figure}[ht!]
    \centering
    \includegraphics[width=.45\textwidth]{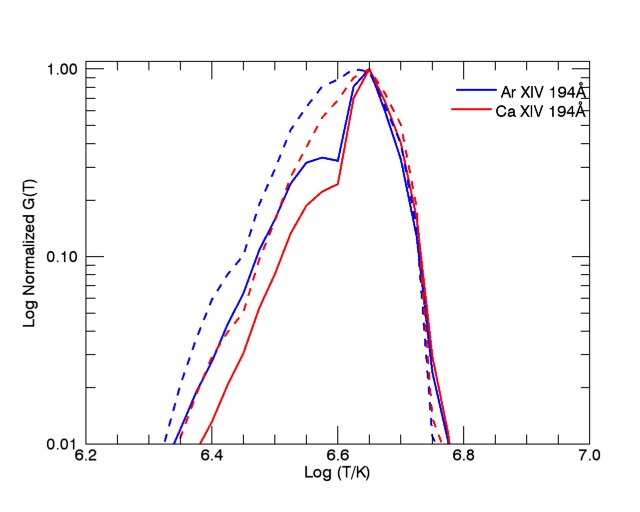}
    \caption{Log normalised contribution functions, G(T) of \ion{Ca}{14}~193.87~\AA and \ion{Ar}{14}~194.40\AA\ convolved with the DEM for the flaring region. Dashed lines represent DEMs calculated from the 12:29~UT raster; Solid lines were calculated using the 14:19~UT (flaring) raster.}
    \label{fig:SiS_FIP_hist}
\end{figure}

In this section, we address the concern of temperature effect on the \ion{Ca}{14}/\ion{Ar}{14} intensity ratio value. In Figure~\ref{fig:gofnt}, the electron temperature at the flaring region ranges from $\mathrm{log_{10}6.59-log_{10}6.68}$. This puts the theoretical intensity ratio at the uphill region in panel a of Figure~\ref{fig:gofnt}, suggesting that a change in temperature could produce the observed change in the intensity ratio. A differential emission measure (DEM) was performed around the region of interest. Over the temperature range, two curves coincide very well, and the peak temperature does not change significantly, suggesting the dramatic increase in intensity ratio value is real.

\acknowledgements{
A.S.H.T. thanks the STFC for support via funding given in his PHD studentship. D.M.L. is grateful to the Science Technology and Facilities Council for the award of an Ernest Rutherford Fellowship (ST/R003246/1). The work of D.H.B. was performed under contract to the Naval Research Laboratory and was funded by the NASA Hinode program. D.B. is funded under STFC consolidated grant number ST/S000240/1 and L.v.D.G. is partially funded under the same grant. L.v.D.G. acknowledges the Hungarian National Research, Development and Innovation Office grant OTKA K-113117. J.M.L. was supported by the NASA Heliophysics Guest Investigator (80HQTR19T0029) and Supporting Research (80HQTR20T0076) programs, and by Basic Research Funds of the Office of Naval Research. G.V. acknowledges the support from the European Union's Horizon 2020 research and innovation programme under grant agreement No 824135 and of the STFC grant number ST/T000317/1. Hinode is a Japanese mission developed and launched by ISAS/JAXA, with NAOJ as domestic partner and NASA and STFC (UK) as international partners. It is operated by these agencies in co-operation with ESA and NSC (Norway). AIA data courtesy of NASA/SDO and the AIA, EVE, and HMI science teams. CHIANTI is a collaborative project involving George Mason University, the University of Michigan (USA) and the University of Cambridge (UK).
}

\bibliography{bib}{}
\bibliographystyle{aasjournal}



\end{document}